\documentclass{article}
\usepackage[T1]{fontenc} 
\usepackage[utf8]{inputenc} 
\usepackage{ismir,amsmath,cite,url}
\usepackage{graphicx}
\usepackage{color}
\usepackage{multirow}
\usepackage{booktabs}
\usepackage{dblfloatfix}    
\usepackage{enumitem}  
\let\OLDthebibliography\thebibliography
\renewcommand\thebibliography[1]{
  \OLDthebibliography{#1}
  \setlength{\parskip}{3pt}
} 

\usepackage[acronym]{glossaries}
\newacronym{AFP}{AFP}{audio fingerprinting}
\newacronym{DL}{DL}{deep learning}
\newacronym{STFT}{STFT}{short time Fourier transform}
\newacronym{AE}{AE}{auto-encoder}
\newacronym{ME}{ME}{music enhancement}
\newacronym{DCASE}{DCASE}{Detection and Classification of Acoustic Scenes and Events}
\newacronym{IR}{IR}{impulse response}
\newacronym{dB}{dB}{decibels}
\newacronym{ms}{ms}{milliseconds}
\newacronym{TF}{TF}{time-frequency}
\newacronym{PSNR}{PSNR}{Peak Signal to Noise Ratio}

\usepackage{lineno}

\title{Music Augmentation and Denoising \\ for Peak-Based Audio Fingerprinting}

\threeauthors
  {Kamil Akesbi} {Deezer\\{\tt research@deezer.com}}
  {Dorian Desblancs} {Deezer \\ {\tt research@deezer.com}}
  {Benjamin Martin} {Deezer \\ {\tt research@deezer.com}}

\def\authorname{Kamil Akesbi , Dorian Desblancs, and Benjamin Martin}

\usepackage[bookmarks=false,pdfauthor={\authorname},pdfsubject={\papersubject},hidelinks]{hyperref}
\begin{document}

\maketitle

\begin{abstract}

Audio fingerprinting is a well-established solution for song identification from short recording excerpts. Popular methods rely on the extraction of sparse representations, generally spectral peaks, and have proven to be accurate, fast, and scalable to large collections. However, real-world applications of audio identification often happen in noisy environments, which can cause these systems to fail. In this work, we tackle this problem by introducing and releasing a new audio augmentation pipeline that adds noise to music snippets in a realistic way, by stochastically mimicking real-world scenarios. We then propose and release a deep learning model that removes noisy components from spectrograms in order to improve peak-based fingerprinting systems' accuracy. We show that the addition of our model improves the identification performance of commonly used audio fingerprinting systems, even under noisy conditions.

\end{abstract}  

\section{Introduction}\label{sec:introduction}

Music identification consists of recognizing a recording, e.g. a song in a database, given a music audio extract as input \cite{Sonnleitner2017}.  This extract can come from numerous sources, such as a mobile recording or music mix \cite{sonnleitner2016landmark}.
The technology behind most well-known music identifications apps is called \gls{AFP} \cite{AudioFingerprintingConcepts}. Compact and discriminative audio features, called audio fingerprints, are extracted from a query audio segment and compared to an indexed reference database containing precomputed audio fingerprints as well as their corresponding metadata information \cite{cano2005review}. In the case of a successful identification, the content information linked to the identified fingerprints is retrieved and sent to the user. 

A fingerprint extraction algorithm is usually designed to be deterministic, scalable, and robust to a wide range of audio transformations \cite{RobustAFP}. It can be designed to identify a song from an audio recording in situations where the audio has been compressed, undergone pitch, tempo, or speed transformations, or been recorded in the presence of strong background noise.

One popular category of \gls{AFP} algorithms is based on the concept of spectral peaks, also known as peak-based \gls{AFP}. Many of these algorithms, including the industrial scale \textit{Shazam algorithm} \cite{shazam}, have been developed and improved over the years to handle common audio transformations \cite{cano2005review}. However, most systems still struggle when background noise is present in the recording environment \cite{Sonnleitner2017}. Recent works \cite{NowPlaying, NeuralAudioFingerprint, ContrastiveAFP, SAMAF, AttentionQueryByExample} have proposed \gls{DL} based fingerprint extraction algorithms, sometimes trained to be more robust to background noise. However, these methods often rely on heavy vectorial audio representations, that are difficult to store, index and compare when databases contain millions of audio tracks \cite{Sonnleitner2017, NeuralAudioFingerprint}.

\begin{figure*}[h]
 \centering
 \includegraphics[width=0.85\textwidth]{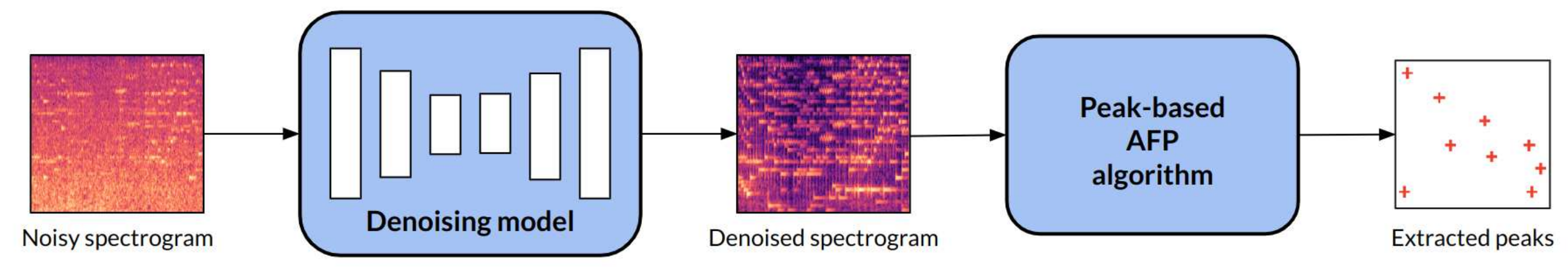}
 \caption{Overview of the proposed approach. First, a noisy spectrogram is used as input to an Encoder-Decoder model trained to denoise musical snippets. This spectrogram can be the result of real-world environments, such as the audio generated by a phone microphone, or simulations, such as the audio generated by our proposed augmentation pipeline. The denoised spectrogram is then used as input to existing peak-based \gls{AFP} systems.}
 \label{fig:Overview}
\end{figure*}

In this work, we propose a new approach in which a \gls{DL} model is inserted in front of a peak-based \gls{AFP} algorithm. To the best of our knowledge, this is the first work that aims to leverage \gls{DL} in order to improve peak-based \gls{AFP} algorithms' robustness to noise present in real environments. The model is trained to remove noise from audio snippets while keeping musical content salient enough to be successfully matched through a traditional \gls{AFP} algorithm. The noisy snippets are generated using a data augmentation pipeline that mimicks the recording of a song played in a noisy environment. \figref{fig:Overview} gives an overview of the proposed approach. 

This hybrid strategy has several advantages.  
First, the complete pipeline not only provides robustness to transformations that are already handled by a peak-based system, such as equalization, dynamic compression, and lossy encoding artifacts, but also to realistic noise addition. Second, the \gls{DL} model can be inserted into any \gls{AFP} system: neither the fingerprints nor the way they are stored need to be changed in order to fit the system. Finally, the scalability and computational efficiency of peak-based \gls{AFP} systems is preserved. The generated fingerprints are very light and can be indexed using fast, classic \gls{AFP} methods \cite{shazam, ReviewAFP}. 

Our main contributions are as follows:
\begin{itemize}[noitemsep,topsep=0pt,parsep=0pt,partopsep=0pt,leftmargin=*]
    \item A strong data augmentation pipeline that can be used to simulate realistic noise on studio-recorded music. We focus on modeling noises that are present in places where music is usually played. 
    \item A \gls{DL} model that can denoise spectrogram representations of mobile phone audio recordings of music, trained using augmented examples. When used in conjunction with a peak-based \gls{AFP} algorithm, the model allows us to identify more augmented songs, most notably due to the greater proportion of preserved spectrogram peaks between clean and denoised audio. 
\end{itemize}

\section{Related Work}\label{sec:RW}

\subsection{Audio Fingerprinting}


Audio fingerprint systems rely on two steps: the extraction of an audio fingerprint from an audio recording and the search for matching fingerprints in a database \cite{RobustAFP, burges2003distortion, seo2006audio, cano2005review, haitsma2003highly, ReviewAFP}. Peak-based \gls{AFP} was introduced in \cite{shazam}. After computing a \gls{STFT} representation of the audio signal, a peak-picking strategy is used to extract \gls{TF} points that have the highest magnitude in a given neighborhood. Peaks with high magnitude are more likely to survive in the presence of additive noise and distortions related to the recording device encoding. However, the robustness of such approaches is limited when a significant amount of background noise is present in the signal \cite{shazam}.
In 2014, \cite{panako} proposed to store the relative position of triplets of peaks as fingerprint hashes to design an algorithm capable of handling time stretching, pitch shifting and time scale modifications of the signal. Another similar approach \cite{quad} proposes an efficient selection and grouping of the spectral peaks in sets of four, called quads. This allows the system to handle time-scale and pitch modifications while increasing the hash specificity compared to its triplet version.

Recently, an increasing amount of \gls{AFP} works have used \gls{DL} methods \cite{NowPlaying, SAMAF, ContrastiveAFP, NeuralAudioFingerprint, AttentionQueryByExample}. These models are usually trained using self-supervised learning strategies with contrastive or triplet losses that generate similar embeddings for similar audio inputs, i.e degraded versions of the same audio segment in our case. In \cite{ContrastiveAFP, NeuralAudioFingerprint}, the authors employ data augmentation techniques to train their models. 
Although these models have better performances than traditional \gls{AFP} methods, they have mostly been studied on relatively small datasets ($\sim$10k tracks for \cite{NowPlaying}, $\sim$100k tracks in \cite{NeuralAudioFingerprint}, $\sim$4500 hours of audio in \cite{ContrastiveAFP}). In particular, methods for indexing and storing such fingerprints are rarely described, and the way to perform efficient search on such datasets at a commercial scale still has to be addressed. A notable exception is the Google Sound Search system that can process a musical library of over 10 million audio tracks through spatial partitioning, vector quantization and
efficient search strategies \cite{GoogleSoundSearch}.

\subsection{Audio Denoising}\label{sec:AudioDenoising}

Audio denoising, or enhancement, aims to attenuate the noise present in the recorded audio without modifying the original signal. 
Traditional audio denoising approaches are based on signal processing techniques such as spectral substraction \cite{SpecralSubstraction}, Wiener Filtering \cite{WienerFiltering}, and Bayesian estimators \cite{BayesianEstimator}. 
Recently, \gls{DL} approaches have shown significant improvements over traditional methods, in both the waveform and time-frequency domain. Commonly, \gls{AE} architectures are trained in a supervised manner using clean and noisy audio pairs \cite{DenoisingAE, DenoisingAE2, LSTM}. 
The U-Net architecture \cite{U-net} is a powerful \gls{AE} model to perform audio enhancement. 
It has been widely studied and used in the fields of audio denoising, speech enhancement, and music source separation due to its state-of-the-art performance in numerous works\cite{WaveUnet,Demucs, Segan, AeGAN, LearningToDenoiseHistoricalMusic, TwoStageUnet, DemucsMusicSeparation,defossez2019music}. Even though other approaches, such as attention-based models, have outperformed U-Net architectures on speech enhancement, audio denoising and source separation tasks \cite{SepFormer, SETransformer, TSTTN, CMGAN, DPT-FSNet, MANNER, AIAT}, U-Net architectures are still largely used in music analysis due to their lightness and efficiency, especially for applications involving real-time processing \cite{Spleeter, UnetMusicSourceSeparation}. 

Music denoising has not been studied as thoroughly as speech enhancement. Previous works studied the restoration of old analog discs \cite{TwoStageUnetHistRec, LearningToDenoiseHistoricalMusic}, which can have poor quality due to hiss, clicks, or scratches in the signal \cite{AudioRestoration}. More recently, \cite{MusicEnhancement} studies how low-quality music recordings, such as the ones captured by mobile devices, can be enhanced using \gls{DL}. In particular, it proposes a realistic data degradation pipeline that greatly inspired this work. Other musical audio augmentations frameworks were already proposed for instance in \cite{McFee2015ASF}, but these are generally targeted towards very different goals, such as creating perturbations in annotated musical datasets in that case.

\section{Music augmentation}

\begin{figure*}[h!]
 \centering
 \includegraphics[width=0.85\textwidth]{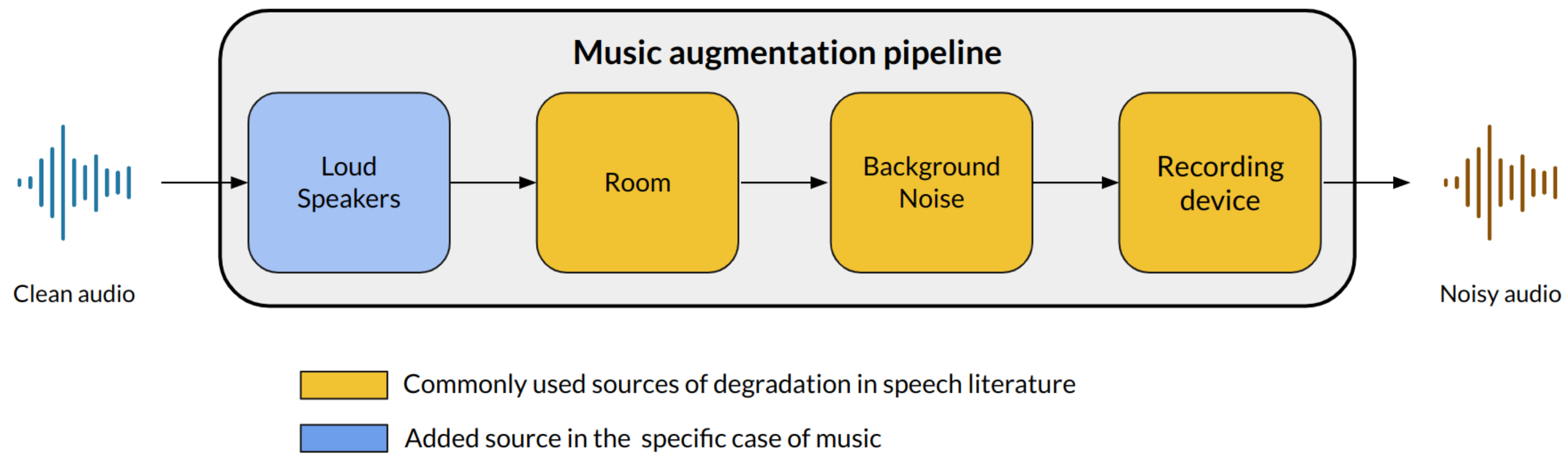}
 \caption{Overview of our proposed music augmentation pipeline. Original audio masters are passed through four augmentation layers to generate noisy audio segments. The transformations in each layer are applied sequentially in order to mimick a sound being played by a speaker, that then travels through an environment full of other noises, before arriving to a recording device. The simulated recorded audio then needs to be identified.}
 \label{fig:SourcesOfNoise}
\end{figure*}


In order to train \gls{DL} models for \gls{AFP} in noisy environment, we introduce a novel, realistic audio augmentation pipeline to create clean-noisy pairs. The augmentation pipeline is composed of several transformations applied to an audio input. These aim to simulate the degradations that recordings may undergo in places where music identification apps are used.  Each transformation models a real-world alteration of the recorded audio, and either comes from the device producing the sound, the surrounding environment, or the recording device itself. In particular, our augmentation pipeline does not include  pitch, speed, or tempo transformations, since some peak-based \gls{AFP} systems already handle these transformations \cite{panako,quad}. When creating our audio augmentation system, we aimed to reproduce degradations caused by room responses, background noise, recording devices, and loud speakers. The first three degradations are commonly found in the speech and audio denoising literature \cite{borsos2021micaugment, DataAugmentationThesis} to model the surrounding noise of people talking, for example.

In the case of music, the reference fingerprints correspond to features extracted from high-quality original recordings. On top of room responses, background noise, and recording devices, loudspeakers playing music in bars, cafes or cars, for instance, can significantly alter the music quality. We consider them as a fourth source of degradation. 
We apply the selected transformations sequentially to 44.1 kHz raw audio waveforms, using the Python library \textit{torch-audiomentations}\footnote{\url{https://github.com/asteroid-team/torch-audiomentations}}. We provide the code to generate augmented content, as well as audio examples\footnote{\url{https://github.com/deezer/MusicFPAugment}}.
The following sections provide a detailed overview of each augmentation layer. 

\subsection{Background Noise}

In order to model non-stationary, additive noise that intervenes in places where music is typically played, we mix musical snippets with samples coming from past \gls{DCASE} challenges. These challenges occur every year and present new or expanded acoustic scene datasets. These scenes are recorded in different places and environments with different devices \cite{DCASE_challenge_2018, DCASE_challenge_2020}.
We build our own background noise dataset using samples from the 2017 TUT Acoustic scenes challenge \cite{Mesaros2016_EUSIPCO}, 2018 TUT Urban Acoustic Scenes challenge \cite{Mesaros2018_DCASE}, and 2020 TAU Urban Acoustic Scenes 2020 Mobile challenge \cite{Mesaros2018_DCASE}.

\begin{table}[h!]
\centering
\resizebox{0.9\columnwidth}{!}{
\begin{tabular}{|c c c c|}
\hline
Airport & City center  & Mall  & Residential area \\
Beach & Forest & Metro & Street pedestrian \\
Bus & Grocery store & Office & Street traffic \\
Café & Home & Park & Train \\
Car & Library & Public square & Tram \\
\hline 
\end{tabular}
 }
\caption{Background noise dataset acoustic scene classes.}
\label{tab:NoiseClasses}
\end{table}

Combining audio from these datasets leads to an unbalanced dataset containing 56,340 10-second samples from 20 different classes (see \tabref{tab:NoiseClasses}). 
We however randomly select 441 samples per class, for a total of 8,820 samples, in order to create a balanced background noise dataset to train and validate our denoising model.
Note that we use a 90-10 train and validation split over each class.
Samples from the remaining 47,520 items are then randomly selected for testing.
When mixing the samples with our musical snippets, we randomly select a signal-to-noise ratio in the range of $[-10,10]$ \gls{dB} in order to mimick a variety of different realistic settings.

\begin{figure*}[h!]
 \centering
 \includegraphics[width=\textwidth]{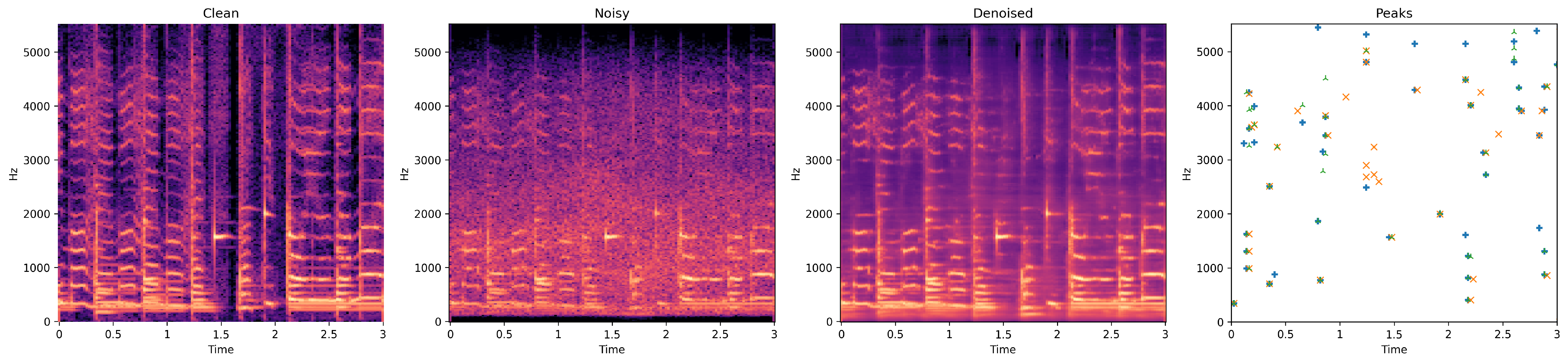}
 \caption{Example of clean, noisy and denoised spectrograms (in \gls{dB}) obtained from a three-second music snippet. The last plot shows the spectral peaks extracted from each spectrogram using one AFP system: Audfprint. Blue peaks come from the clean spectrogram, orange peaks from the noisy one, and green peaks from the denoised one.}
 \label{fig:ModelBehaviorAnalysis}
\end{figure*}

\subsection{Room Impulse Response}

We select the MIT \gls{IR} survey dataset \cite{MIT_IR_survey}, which consists of 271 real impulse responses sampled at 32 kHz measured in a large variety of distinct locations, to simulate environmental reverberation. 
Due to the small nature of this dataset, which still contains a great variety of \gls{IR}s, we did not split this dataset into train, val and test sets.
Note that, although other datasets contain a broader range of \gls{IR}s\cite{rir-datasets}, we judged this one to better balance the tradeoff between size, variety and realism.

\subsection{Recording Device}

In industrial applications, audio recordings may be performed by a large range of microphones, often embedded in devices of varying quality. We model the degradations induced by the recording device by adding:  \textbf{1.~Gain}: during the recording, signal volume can vary. We model this phenomenon by multiplying the audio by a random factor between $[-5,5]$ \gls{dB}. We apply this transformation with a probability of 0.3 to the audio samples in our training set. \textbf{2.~Clipping}: samples with high magnitude can be clipped during recording. We model this by artificially clipping all samples above a certain threshold. The threshold is selected randomly so that between 0 and 1\% of the audio samples with highest magnitude within each snippet are clipped. \textbf{3.~Low-pass and high-pass filters}: unfortunately, industry makers very rarely share public information about frequency response for embedded microphones. Therefore, in the absence of a comprehensive review on smartphone microphone responses, we choose to approximate distortions by first applying a low-pass filter with a randomly selected cutoff frequency in the range $[3500,7000]$ Hz and a first order high-pass filter with cutoff frequency in the range $[30, 150]$ Hz.

As one can notice, all these transformations are applied with a certain probability or using a randomly selected value in a certain range.
This was done on purpose; in order to increase the variability of each augmented audio while keeping it in a zone we judged realistic, that simulated the real world well.

\subsection{Loudspeakers}

As explained above, sound reproduction quality is likely to vary in practical recordings depending on the loudspeaker.
Unfortunately, similarly to microphone devices, it is hard to get a comprehensive overview of speaker frequency responses in practice. 
In an attempt to simulate such degradations, we extract the -3\gls{dB} cutoff frequency of 431 loudspeakers from the Spinorama website \cite{spinorama}. This website contains statistics about loudspeakers, and displays that most cutoff frequencies are located between 20 and 150 Hz. 
We choose to model the speakers' frequency responses using first-order high-pass filters with cutoff frequencies in a similar range. For both microphone and loudspeaker simulation, we acknowledge that approximations might limit the realism of the generated data, and leave more sophisticated modelling as future work. We would like to be able to take into account non-linear distortions, for instance.

 

\section{Audio Denoising Model}

\subsection{Model Selection and Training}

\begin{table*}[h!]
\centering
\resizebox{0.85\textwidth}{!}{\begin{tabular}{@{}cccccccccccc@{}}
\toprule
\multicolumn{1}{l}{}                                                          & \multicolumn{1}{l}{}         & \multicolumn{5}{c}{\textbf{Audfprint}}                                                          & \multicolumn{5}{c}{\textbf{Dejavu}}                                                                                                         \\ \midrule
\multicolumn{2}{c|}{\textbf{Configuration}}                                                                  & \multicolumn{2}{c}{\textbf{Denoising}}                  & \multicolumn{3}{c|}{\textbf{Peak preservation}}                                          & \multicolumn{2}{c}{\textbf{Denoising}}                  & \multicolumn{3}{c}{\textbf{Peak preservation}}                                    \\ \midrule
Set up                                                                        & \multicolumn{1}{c|}{Dataset} & \textbf{L1 loss $\downarrow$}           & \textbf{PSNR  $\uparrow$}              & \textbf{Prec $\uparrow$}             & \textbf{Rec $\uparrow$}              & \multicolumn{1}{c|}{\textbf{F1 $\uparrow$}} & \textbf{L1 loss $\downarrow$}           & \textbf{PSNR $\uparrow$}              & \textbf{Prec $\uparrow$}             & \textbf{Rec $\uparrow$}              & \textbf{F1 $\uparrow$}               \\ \midrule
\multirow{3}{*}{\textbf{\begin{tabular}[c]{@{}c@{}}Our\\ Model\end{tabular}}} & \multicolumn{1}{c|}{Train}   & \textbf{0.0103}                     & 29.838                     & 0.363                     & 0.305                     & \multicolumn{1}{c|}{0.302}       & \textbf{0.0105}                     & 34.130                      & 0.351                     & 0.305                     & 0.326                     \\ \cmidrule(l){2-12} 
             & \multicolumn{1}{c|}{Val}     & 0.0106                     & 29.856                     & \textbf{0.367}                     & \textbf{0.307}                     & \multicolumn{1}{c|}{\textbf{0.333}}       & 0.0110                     & 34.250                     & \textbf{0.321}                     & \textbf{0.320}                     & \textbf{0.320}                     \\ \cmidrule(l){2-12} 
                                                                              & \multicolumn{1}{c|}{Test}    & 0.0106 & \textbf{30.545} & 0.347 & 0.298 & \multicolumn{1}{l|}{0.316}       & 0.0110 & \textbf{35.119} & 0.319 & 0.270 & 0.288 \\ \midrule
\multirow{2}{*}{\textbf{\begin{tabular}[c]{@{}c@{}}No\\ model\end{tabular}}}  & \multicolumn{1}{c|}{Val}     & 0.0180                     & 26.505                     & 0.285                     & 0.239                     & \multicolumn{1}{c|}{0.259}       & 0.0190                     & 31.751                     & 0.272                     & 0.253                     & 0.262                     \\ \cmidrule(l){2-12} 
                                                                              & \multicolumn{1}{c|}{Test}    & 0.0187                     & 27.112 & 0.278 & 0.238 & \multicolumn{1}{l|}{0.253}       & 0.0190 & 32.622 & 0.257 & 0.245 & 0.246 \\ \bottomrule
\end{tabular}}
 \caption{Train, validation, and test set denoising and peak preservation performance. We report the mean results obtained on five metrics (L1 loss, \gls{PSNR}, precision, recall, and F1-score) using two \gls{AFP} systems: Audfprint and Dejavu.}
 \label{tab:Trainings}
\end{table*}

As explained in \ref{sec:AudioDenoising}, U-Net architectures have been shown to perform well on audio denoising tasks. We use and adapt the model from \cite{U-net}, as it is sufficiently fast and light to be used in industrial query-by-example applications. It has a relatively low inference time as it is mostly constituted of convolutional layers. We measured an average inference time on 10-second \gls{STFT}s, using a single CPU with two threads, of 753 \gls{ms}.
It is hence more than ten times faster than real-time using relatively limited resources. 
We experimented with more recent, transformer-based architectures in \cite{akesbi2022audio}, but the simpler and faster U-Net architecture proved to be higher-performing.
We leave the optimization of our model architecture as future work. 

As explained in \cite{MusicEnhancement}, it is easier to denoise music in the \gls{TF} domain since music is polyphonic; additive sources usually cover different regions of the frequency spectrum and can thus be more easily identified using frequency features of the signal. We, therefore, train our model to denoise audio spectrograms. More precisely, we solely focus on the magnitude component of the spectrograms since we are only interested in spectral peak preservation, and do not seek to re-synthesize the denoised audio samples. 
The U-Net is trained to predict the clean spectrograms in a supervised fashion, using an L1 loss, a batch size of 32, the Adam optimizer \cite{kingma2014adam} with a learning rate first set at $0.001$, a learning rate scheduler with patience 2 and a decreasing factor of 0.5. We compute the validation loss every tenth of a training epoch and save the best-performing weights for testing.  We also use a dropout rate $p=0.05$ after each downsampling block of the encoder architecture.

\subsection{Training Set} \label{sec:training-set}

In order to train our model on a wide variety of musical content, we randomly select three-second audio snippets from 48,872 commercial audio tracks that come from expertly curated playlists and cover 25 human-annotated musical genres. 
The dataset is then separated into training, validation, and test sets using an 80-10-10 split for each genre category. We generate a set of five augmented samples at 44.1 kHz from each clean audio song by training the denoising model with different noisy versions of the same clean audio excerpt.
This was found to improve the model's generalization ability.
Our final dataset consists of 244,360 3-second clean-noisy audio pairs. 
Due to computational constraints, augmentations are not performed online during training.

\subsection{Evaluation Strategy}

To evaluate our approach, we propose two complementary series of experiments applied to open-source \gls{AFP} systems. In the first one, presented in section \ref{sec:denoising-results}, we study the behavior of our model in terms of denoising capability and peak preservation. In the second one, presented in section \ref{sec:impact-identification}, we study how inserting the denoising model into a \gls{AFP} system actually impacts its identification performance.

\section{Peak Preservation}\label{sec:denoising-results}

\subsection{Evaluation Framework}

We first select two popular open-source \gls{AFP} systems: Audfprint \cite{ellis14}
and Dejavu \cite{dejavu}. Both are implementations of the conventional industrial method described in~\cite{shazam}. They extract spectral peaks in different ways from \gls{STFT} spectrograms of the audio signal, then group them into couples to form landmarks that may be hashed and later indexed into a dataset for efficient search. These are commonly used as fingerprinting baselines in the literature \cite{NeuralAudioFingerprint, AttentionQueryByExample, ContrastiveAFP}.  

We use a similar setup for both systems. First, audio signals are downsampled to 11,025 Hz and \gls{STFT}s are extracted with a hop size of 256 and a frame size of 512. However, the magnitude spectrograms input to the \gls{AFP} system are computed slightly differently in Audfprint and Dejavu. We, therefore, train a separate model for each system that adequately fits the \gls{AFP} system whose input it is supposed to denoise\footnote{Model weights and experimental code are available in this publication's GitHub repository.}.

We first evaluate the denoising capability of the model, by calculating metrics that compare the noisy and denoised spectrograms that are input to the \gls{AFP} systems. We report the L1 distance, which is used for training, and the \gls{PSNR}, another commonly used metric to quantify image reconstruction quality \cite{hore2010image}.
Additionally, in order to highlight peak preservation, we use metrics that assess the proportion of spectral peaks that are found in both the noisy and clean signals. 
We pass the input spectrograms through peak extraction steps of the \gls{AFP} systems, and then look at whether the coordinates of extracted peaks match. We then report the usual retrieval metrics from the the list of matching peaks: precision, recall, and F1-score.

\begin{table*}[h!]
\centering
\resizebox{0.64\textwidth}{!}{\begin{tabular}{@{}lcccccccc@{}}
\toprule
\multicolumn{3}{c}{\textbf{Configuration}}                                                                            & \multicolumn{3}{c}{\textbf{Audfprint}}                                & \multicolumn{3}{c}{\textbf{Dejavu}}              \\ \midrule
\textbf{Dataset}                                   & \textbf{Queries}       & \multicolumn{1}{c|}{\textbf{Denoising}} & \textbf{3s}    & \textbf{5s}    & \multicolumn{1}{c|}{\textbf{10s}}   & \textbf{3s}    & \textbf{5s}    & \textbf{10s}   \\ \midrule
\multicolumn{1}{c}{\multirow{4}{*}{\textbf{Test}}} & Clean                  & \multicolumn{1}{c|}{No model}           & 89.0           & 96.0           & \multicolumn{1}{c|}{99.0}           & 83.0          & 91.9          & 98.0          \\ \cmidrule(l){2-9} 
\multicolumn{1}{c}{}                               & \multirow{3}{*}{Noisy} & \multicolumn{1}{c|}{No model}           & 43.3          & 61.8          & \multicolumn{1}{c|}{79.3}          & 30.4          & 37.6          & 57.5          \\
\multicolumn{1}{c}{}                               &                        & \multicolumn{1}{c|}{Our model}          & 39.7          & 61.2          & \multicolumn{1}{c|}{80.9}          & 30.2          & 38.2          & 59.6          \\
\multicolumn{1}{c}{}                               &                        & \multicolumn{1}{c|}{Mix}                & \textbf{48.9} & \textbf{67.3} & \multicolumn{1}{c|}{\textbf{84.1}} & \textbf{38.8} & \textbf{48.0} & \textbf{68.0} \\ \midrule
\multirow{4}{*}{\textbf{FMA large}}                & Clean                  & \multicolumn{1}{c|}{No model}           & 80.8          & 93.9          & \multicolumn{1}{c|}{96.8}          & 81.5          & 91.2          & 96.7          \\ \cmidrule(l){2-9} 
                                                   & \multirow{3}{*}{Noisy} & \multicolumn{1}{c|}{No model}           & 24.6          & 44.8          & \multicolumn{1}{c|}{64.6}          & 21.8          & 32.2          & 49.1          \\
                                                   &                        & \multicolumn{1}{c|}{Our model}          & 25.5          & 48.3          & \multicolumn{1}{c|}{71.4}          & 19.5          & 29.4          & 46.5          \\
                                                   &                        & \multicolumn{1}{c|}{Mix}                & \textbf{31.9} & \textbf{53.8} & \multicolumn{1}{c|}{\textbf{74.0}}  & \textbf{26.0} & \textbf{38.2} & \textbf{55.9} \\ \bottomrule
\end{tabular}}
\caption{Music identification rate for each AFP system. We report results on three, five, and ten-second clean and noisy queries from our test dataset and the FMA large dataset. Noisy query results are reported with and without the \gls{DL} model, as well as in a \textit{Mix} setting where hashes from both denoised and noisy audio are used to identify the input snippet.}
\label{tab:Test}
\end{table*}

\subsection{Results}

\tabref{tab:Trainings} summarizes the mean results obtained for both \gls{AFP} systems. Again, each system computes spectrograms and extracts peaks in a different way, which explains why metrics are different from one \gls{AFP} system to another. 

First, our experiments yield similar results on the train, validation, and test subsets of our augmented dataset. This suggests that our model does not overfit training data.
Moreover, we improve upon all metrics using our denoising model on the validation and test sets compared to the no-model baseline. Each model is able to reduce the amount of noise present in noisy spectrograms and make them closer to their clean references. The L1 loss is accordingly decreased while \gls{PSNR} is increased on the test set, going from 27.112 to 30.545 on Audfprint and 32.622 to 35.119 on Dejavu. When looking at the output denoised spectrogram (see \figref{fig:ModelBehaviorAnalysis} for an example), we observe that the model indeed smooths out noisy content while keeping most of the musical components.

The denoising models also help each \gls{AFP} system preserve the spectral peaks that are present in the clean audio. The F1-score obtained improves using both Audfprint and Dejavu; both systems are able to remove peaks induced by noise addition, as shown by the precision increase of 6.9\% for Audfprint and 6.2\% for Dejavu on the augmented test set, for instance. Additionally, our model also allows us to restore peaks masked by signal degradations, as shown by the recall increase of 6.0\% for Audfprint and 2.5\% for Dejavu on this same test set.

  

\section{Music Identification}\label{sec:impact-identification}


\subsection{Evaluation Framework}

In order to test Audfprint and Dejavu retrieval systems, we first need to build an indexed dataset. 
As explained in \ref{sec:training-set}, our test set already contains 4,888 audio tracks that we use as a first, reference dataset. However, in order to conduct tests on a larger, more realistic catalog, we build a second reference dataset with tracks coming from the FMA large (Free Music Archive) dataset \cite{fma_dataset}. After removing corrupted files, the dataset contains 105,721 high-quality tracks that span 161 different musical genres.

For each reference dataset, we randomly extract audio snippets of 3, 5 and 10 seconds from different tracks. These correspond to our clean queries. We then process them with the music augmentation pipeline to create noisy queries. For each snippet size, we generate 4,888 clean and as many noisy queries for our test set, and 10,000 clean and as many noisy queries for the FMA set.
We evaluate the \gls{AFP} systems with the common \cite{NeuralAudioFingerprint,shazam} top-1 hit identification rate metric, defined as:

\begin{equation*}
    \textit{Id. Rate} = \frac{\textit{n  hits @ top-1}}{\textit{n hits @ top-1 + n miss @ top-1}} * 100 
\end{equation*}

\subsection{Results}

The results obtained are summarized in \tabref{tab:Test}. We observe that the identification rate significantly drops for noisy queries.
On the FMA large dataset, with Audfprint, the identification rate decreases from 80.8\% for clean queries to 24.6\% for noisy queries in a three second setting.
With Dejavu, a similar identification rate drop is observed, that goes from 81.5\% for clean queries to 21.8\% for noisy queries.
These results display the limitations of peak-based \gls{AFP} systems when they are exposed to noisy environments and highlight the need to improve their performance.

Passing the noisy content through our denoising model before identification does not yield results that significantly differ from a no-model setting. The performance with denoising is either slightly higher or slightly lower than the fully-noisy setting depending on the query size, database and AFP system. For example, on our test set, in a five second setting, the noisy identification rate is 61.8\% and the denoised identification rate is 61.2\%. Dejavu displays values of 37.6\% and 38.2\% in the same context.

However, we notice that a high number of tracks are identified with denoising but not without, and vice versa. Combining the two approaches in a \textit{Mix} pipeline appeared to significantly outperform baselines. We use a dual-query setup where both clean and denoised queries are passed to the identification task. Each query returns an identified track when appropriate. In the case where identified tracks differ, we rank them using the same method Audfprint and Dejavu use to rank identified songs, by considering that the track with the highest number of common hashes aligned in time with the query is the correct one. This approach allows us to obtain significantly better results. In a ten second setting, on our test set, the identification rate increases by 4.8\% for Audfprint and 10.5\% for Dejavu compared to the noisy query setting. On the FMA dataset, in a three second setting, the identification rate increases by 7.3\% for Audfprint and 4.2\% for Dejavu.

Looking at how peaks are reconstructed or preserved gives us an intuition about why the denoising model helps the systems in some cases and penalizes it in others. In \figref{fig:ModelBehaviorAnalysis}, the denoising model helps remove noise, and thus incorrect peaks, from regions where an important amount of noise is added: for example, on the last plot, orange points that do not overlap with any blue point are noisy peaks that are correctly removed by the denoising model. However, the denoising model can also cause the AFP system to extract incorrect peaks (green points that do not overlap with any blue point). These incorrectly predicted peaks partly explain why the AFP system performs better without denoising in some cases.

\section{Conclusion}

 In this paper, we propose to add a music denoising model to an audio fingerprinting system to increase its robustness to realistic noises. Using a new augmentation pipeline, we train a model that can effectively denoise spectrograms and help AFP systems preserve spectral peaks. We find that the model helps the system identify tracks in some cases, but can also penalize it in others. However, combining approaches improves the identification rate by 5 to 10\% in all cases. Future work will mainly focus on better understanding the behavior of the model regarding spectral peak preservation, and studying more complex architectures that learn how to preserve spectral peaks directly.



\newpage
\clearpage
\bibliography{ISMIRtemplate}

\begin{thebibliography}{10}
\providecommand{\url}[1]{#1}
\csname url@samestyle\endcsname
\providecommand{\newblock}{\relax}
\providecommand{\bibinfo}[2]{#2}
\providecommand{\BIBentrySTDinterwordspacing}{\spaceskip=0pt\relax}
\providecommand{\BIBentryALTinterwordstretchfactor}{4}
\providecommand{\BIBentryALTinterwordspacing}{\spaceskip=\fontdimen2\font plus
\BIBentryALTinterwordstretchfactor\fontdimen3\font minus
  \fontdimen4\font\relax}
\providecommand{\BIBforeignlanguage}[2]{{%
\expandafter\ifx\csname l@#1\endcsname\relax
\typeout{** WARNING: IEEEtran.bst: No hyphenation pattern has been}%
\typeout{** loaded for the language `#1'. Using the pattern for}%
\typeout{** the default language instead.}%
\else
\language=\csname l@#1\endcsname
\fi
#2}}
\providecommand{\BIBdecl}{\relax}
\BIBdecl

\bibitem{Sonnleitner2017}
R.~Sonnleitner, ``Audio identification via fingerprinting: Achieving robustness
  to severe signal modifications,'' Ph.D. dissertation, Universit{\"a}t Linz,
  2017.

\bibitem{sonnleitner2016landmark}
R.~Sonnleitner, A.~Arzt, and G.~Widmer, ``Landmark-based audio fingerprinting
  for dj mix monitoring.'' in \emph{Proc. ISMIR}, 2016, pp. 185--191.

\bibitem{AudioFingerprintingConcepts}
P.~Cano, E.~Batlle, E.~G{\'o}mez, L.~de~CT~Gomes, and M.~Bonnet, ``Audio
  fingerprinting: Concepts and applications,'' \emph{Computational Intelligence
  for Modelling and Prediction}, pp. 233--245, 2005.

\bibitem{cano2005review}
P.~Cano, E.~Batlle, T.~Kalker, and J.~Haitsma, ``A review of audio
  fingerprinting,'' \emph{Journal of VLSI signal processing systems for signal,
  image and video technology}, vol.~41, pp. 271--284, 2005.

\bibitem{RobustAFP}
J.~Haitsma and T.~Kalker, ``A highly robust audio fingerprinting system,'' in
  \emph{Proc. ISMIR}, 2002, pp. 107--115.

\bibitem{shazam}
A.~Wang, ``An industrial strength audio search algorithm,'' in \emph{Proc.
  ISMIR}, vol. 2003, 2003, pp. 7--13.

\bibitem{NowPlaying}
B.~Gfeller, B.~Aguera-Arcas, D.~Roblek, J.~D. Lyon, J.~J. Odell, K.~Kilgour,
  M.~Ritter, M.~Sharifi, M.~Velimirović, R.~Guo, and S.~Kumar, ``Now playing:
  Continuous low-power music recognition,'' in \emph{NeurIPS MLPC 2 workshop},
  2017.

\bibitem{NeuralAudioFingerprint}
S.~Chang, D.~Lee, J.~Park, H.~Lim, K.~Lee, K.~Ko, and Y.~Han, ``Neural audio
  fingerprint for high-specific audio retrieval based on contrastive
  learning,'' in \emph{Proc. IEEE ICASSP}.\hskip 1em plus 0.5em minus
  0.4em\relax IEEE, 2021, pp. 3025--3029.

\bibitem{ContrastiveAFP}
Z.~Yu, X.~Du, B.~Zhu, and Z.~Ma, ``Contrastive unsupervised learning for audio
  fingerprinting,'' \emph{arXiv preprint arXiv:2010.13540}, 2020.

\bibitem{SAMAF}
A.~B{\'a}ez-Su{\'a}rez, N.~Shah, J.~A. Nolazco-Flores, S.-H.~S. Huang,
  O.~Gnawali, and W.~Shi, ``Samaf: Sequence-to-sequence autoencoder model for
  audio fingerprinting,'' \emph{ACM TOMM}, vol. 16-2, no.~2, pp. 1--23, 2020.

\bibitem{AttentionQueryByExample}
A.~Singh, K.~Demuynck, and V.~Arora, ``Attention-based audio embeddings for
  query-by-example,'' \emph{arXiv preprint arXiv:2210.08624}, 2022.

\bibitem{ReviewAFP}
P.~Cano, E.~Batle, T.~Kalker, and J.~Haitsma, ``A review of algorithms for
  audio fingerprinting,'' in \emph{IEEE Workshop on Multimedia Signal
  Processing}.\hskip 1em plus 0.5em minus 0.4em\relax IEEE, 2002, pp. 169--173.

\bibitem{burges2003distortion}
C.~J. Burges, J.~C. Platt, and S.~Jana, ``Distortion discriminant analysis for
  audio fingerprinting,'' \emph{IEEE Trans. on Speech and Audio Processing},
  vol.~11, no.~3, pp. 165--174, 2003.

\bibitem{seo2006audio}
J.~S. Seo, M.~Jin, S.~Lee, D.~Jang, S.~Lee, and C.~D. Yoo, ``Audio
  fingerprinting based on normalized spectral subband moments,'' \emph{IEEE
  Signal Processing Letters}, vol.~13, no.~4, pp. 209--212, 2006.

\bibitem{haitsma2003highly}
J.~Haitsma and T.~Kalker, ``A highly robust audio fingerprinting system with an
  efficient search strategy,'' \emph{JNMR}, vol.~32, no.~2, pp. 211--221, 2003.

\bibitem{panako}
J.~Six and M.~Leman, ``Panako: a scalable acoustic fingerprinting system
  handling time-scale and pitch modification,'' in \emph{Proc. ISMIR}, 2014.

\bibitem{quad}
R.~Sonnleitner and G.~Widmer, ``Robust quad-based audio fingerprinting,''
  \emph{IEEE/ACM Trans. Audio. Speech. Lang. Process.}, vol.~24, no.~3, pp.
  409--421, 2015.

\bibitem{GoogleSoundSearch}
J.~Lyon, ``Google’s next generation music recognition blog,''
  \href{https://ai.googleblog.com/2018/09/googles-next-generation-music.html}{blog
  link}, 2018.

\bibitem{SpecralSubstraction}
S.~Boll, ``Suppression of acoustic noise in speech using spectral
  subtraction,'' \emph{IEEE Trans. on Acoustics, Speech, and Signal
  Processing}, vol.~27, no.~2, pp. 113--120, 1979.

\bibitem{WienerFiltering}
P.~Scalart and J.~V. Filho, ``Speech enhancement based on a priori signal to
  noise estimation,'' in \emph{Proc. IEEE ICASSP}, vol.~2, 1996, pp. 629--632.

\bibitem{BayesianEstimator}
P.~C. Loizou, ``Speech enhancement based on perceptually motivated bayesian
  estimators of the magnitude spectrum,'' \emph{IEEE Trans. on Speech and Audio
  Processing}, vol.~13, no.~5, pp. 857--869, 2005.

\bibitem{DenoisingAE}
X.~Lu, Y.~Tsao, S.~Matsuda, and C.~Hori, ``Speech enhancement based on deep
  denoising autoencoder,'' in \emph{Proc. INTERSPEECH}, vol. 2013, 2013, pp.
  436--440.

\bibitem{DenoisingAE2}
X.~Feng, Y.~Zhang, and J.~Glass, ``Speech feature denoising and dereverberation
  via deep autoencoders for noisy reverberant speech recognition,'' in
  \emph{Proc. IEEE ICASSP}, 2014, pp. 1759--1763.

\bibitem{LSTM}
F.~Weninger, H.~Erdogan, S.~Watanabe, E.~Vincent, J.~Le~Roux, J.~R. Hershey,
  and B.~Schuller, ``Speech enhancement with lstm recurrent neural networks and
  its application to noise-robust asr,'' in \emph{Proc. Springer LVA/ICA},
  2015, pp. 91--99.

\bibitem{U-net}
O.~Ronneberger, P.~Fischer, and T.~Brox, ``U-net: Convolutional networks for
  biomedical image segmentation,'' in \emph{Proc. Springer MICCAI}, Munich,
  Germany, 2015, pp. 234--241.

\bibitem{WaveUnet}
C.~Macartney and T.~Weyde, ``Improved speech enhancement with the wave-u-net,''
  \emph{arXiv preprint arXiv:1811.11307}, 2018.

\bibitem{Demucs}
A.~D{\'e}fossez, G.~Synnaeve, and Y.~Adi, ``Real time speech enhancement in the
  waveform domain,'' \emph{arXiv preprint arXiv:2006.12847}, 2020.

\bibitem{Segan}
S.~Pascual, A.~Bonafonte, and J.~Serra, ``Segan: Speech enhancement generative
  adversarial network,'' \emph{arXiv preprint arXiv:1703.09452}, 2017.

\bibitem{AeGAN}
S.~Abdulatif, K.~Armanious, K.~Guirguis, J.~T. Sajeev, and B.~Yang, ``Aegan:
  Time-frequency speech denoising via generative adversarial networks,'' in
  \emph{Proc. IEEE EUSIPCO}, 2021, pp. 451--455.

\bibitem{LearningToDenoiseHistoricalMusic}
Y.~Li, B.~Gfeller, M.~Tagliasacchi, and D.~Roblek, ``Learning to denoise
  historical music,'' \emph{arXiv preprint arXiv:2008.02027}, 2020.

\bibitem{TwoStageUnet}
E.~Moliner and V.~V{\"a}lim{\"a}ki, ``A two-stage u-net for high-fidelity
  denoising of historical recordings,'' in \emph{Proc. IEEE ICASSP}, 2022, pp.
  841--845.

\bibitem{DemucsMusicSeparation}
A.~D{\'e}fossez, N.~Usunier, L.~Bottou, and F.~Bach, ``Demucs: Deep extractor
  for music sources with extra unlabeled data remixed,'' \emph{arXiv preprint
  arXiv:1909.01174}, 2019.

\bibitem{defossez2019music}
------, ``Music source separation in the waveform domain,'' \emph{arXiv
  preprint arXiv:1911.13254}, 2019.

\bibitem{SepFormer}
C.~Subakan, M.~Ravanelli, S.~Cornell, F.~Grondin, and M.~Bronzi, ``On using
  transformers for speech-separation,'' \emph{arXiv preprint arXiv:2202.02884},
  2022.

\bibitem{SETransformer}
W.~Yu, J.~Zhou, H.~Wang, and L.~Tao, ``Setransformer: Speech enhancement
  transformer,'' \emph{Cognitive Computation}, pp. 1--7, 2022.

\bibitem{TSTTN}
K.~Wang, B.~He, and W.-P. Zhu, ``Tstnn: Two-stage transformer based neural
  network for speech enhancement in the time domain,'' in \emph{Proc. IEEE
  ICASSP}.\hskip 1em plus 0.5em minus 0.4em\relax IEEE, 2021, pp. 7098--7102.

\bibitem{CMGAN}
R.~Cao, S.~Abdulatif, and B.~Yang, ``{CMGAN: Conformer-based Metric GAN for
  Speech Enhancement},'' in \emph{Proc. INTERSPEECH}, 2022, pp. 936--940.

\bibitem{DPT-FSNet}
F.~Dang, H.~Chen, and P.~Zhang, ``Dpt-fsnet: Dual-path transformer based
  full-band and sub-band fusion network for speech enhancement,'' in
  \emph{Proc. IEEE ICASSP}, 2022, pp. 6857--6861.

\bibitem{MANNER}
H.~J. Park, B.~H. Kang, W.~Shin, J.~S. Kim, and S.~W. Han, ``Manner: Multi-view
  attention network for noise erasure,'' in \emph{Proc. IEEE ICASSP}, 2022, pp.
  7842--7846.

\bibitem{AIAT}
G.~Yu, A.~Li, C.~Zheng, Y.~Guo, Y.~Wang, and H.~Wang, ``Dual-branch
  attention-in-attention transformer for single-channel speech enhancement,''
  in \emph{Proc. IEEE ICASSP}, 2022, pp. 7847--7851.

\bibitem{Spleeter}
R.~Hennequin, A.~Khlif, F.~Voituret, and M.~Moussallam, ``Spleeter: a fast and
  efficient music source separation tool with pre-trained models,''
  \emph{Journal of Open Source Software}, vol.~5, no.~50, p. 2154, 2020.

\bibitem{UnetMusicSourceSeparation}
V.~S. Kadandale, J.~F. Montesinos, G.~Haro, and E.~G{\'o}mez, ``Multi-channel
  u-net for music source separation,'' in \emph{Proc. IEEE MMSP}, 2020, pp.
  1--6.

\bibitem{TwoStageUnetHistRec}
E.~Moliner and V.~V{\"a}lim{\"a}ki, ``A two-stage u-net for high-fidelity
  denoising of historical recordings,'' in \emph{Proc. IEEE ICASSP}.\hskip 1em
  plus 0.5em minus 0.4em\relax IEEE, 2022, pp. 841--845.

\bibitem{AudioRestoration}
D.~Havelock, S.~Kuwano, and M.~Vorl{\"a}nder, \emph{Handbook of Signal
  Processing in Acoustics}, 2008, vol.~1.

\bibitem{MusicEnhancement}
N.~Kandpal, O.~Nieto, and Z.~Jin, ``Music enhancement via image translation and
  vocoding,'' in \emph{Proc. IEEE ICASSP}, 2022, pp. 3124--3128.

\bibitem{McFee2015ASF}
B.~McFee, E.~J. Humphrey, and J.~P. Bello, ``A software framework for musical
  data augmentation,'' in \emph{Proc. ISMIR}, 2015.

\bibitem{borsos2021micaugment}
Z.~Borsos, Y.~Li, B.~Gfeller, and M.~Tagliasacchi, ``Micaugment: One-shot
  microphone style transfer,'' in \emph{Proc. IEEE ICASSP}, 2021, pp.
  3400--3404.

\bibitem{DataAugmentationThesis}
V.-V. Eklund, ``Data augmentation techniques for robust audio analysis,''
  Master's thesis, 2019.

\bibitem{DCASE_challenge_2018}
A.~Mesaros, T.~Heittola, and T.~Virtanen, ``A multi-device dataset for urban
  acoustic scene classification,'' in \emph{Proc. DCASE}, 2018, pp. 9--13.

\bibitem{DCASE_challenge_2020}
T.~Heittola, A.~Mesaros, and T.~Virtanen, ``Acoustic scene classification in
  dcase 2020 challenge: Generalization across devices and low complexity
  solutions,'' in \emph{Proc. DCASE}, 2020, pp. 56--60.

\bibitem{Mesaros2016_EUSIPCO}
A.~Mesaros, T.~Heittola, and T.~Virtanen, ``{TUT} database for acoustic scene
  classification and sound event detection,'' in \emph{Proc. EUSIPCO}, 2016.

\bibitem{Mesaros2018_DCASE}
------, ``A multi-device dataset for urban acoustic scene classification,'' in
  \emph{Proc. DCASE}, November 2018, pp. 9--13.

\bibitem{MIT_IR_survey}
J.~Traer and J.~H. McDermott, ``Statistics of natural reverberation enable
  perceptual separation of sound and space,'' \emph{Proc. of the National
  Academy of Sciences}, vol. 113, no.~48, pp. E7856--E7865, 2016.

\bibitem{rir-datasets}
``List of room impulse response datasets,''
  \url{https://github.com/RoyJames/room-impulse-responses}, accessed:
  2023-04-13.

\bibitem{spinorama}
``Spinorama : a library to display speaker frequency response and similar
  graphs,'' \url{https://github.com/pierreaubert/spinorama}, accessed:
  2023-04-13.

\bibitem{akesbi2022audio}
K.~Akesbi, ``Audio denoising for robust audio fingerprinting,'' Master's
  thesis, 2022.

\bibitem{kingma2014adam}
D.~P. Kingma and J.~Ba, ``Adam: A method for stochastic optimization,''
  \emph{arXiv preprint arXiv:1412.6980}, 2014.

\bibitem{ellis14}
D.~Ellis, ``The 2014 labrosa audio fingerprint system,'' in \emph{Proc. ISMIR},
  2014.

\bibitem{dejavu}
W.~Drevo, ``Dejavu: open-source audio fingerprinting project,''
  \url{https://github.com/worldveil/dejavu}, accessed: 2023-04-13.

\bibitem{hore2010image}
A.~Hore and D.~Ziou, ``Image quality metrics: Psnr vs. ssim,'' in \emph{Proc.
  IEEE ICPR}, 2010, pp. 2366--2369.

\bibitem{fma_dataset}
M.~Defferrard, K.~Benzi, P.~Vandergheynst, and X.~Bresson, ``{FMA}: A dataset
  for music analysis,'' in \emph{Proc. ISMIR}, 2017.

\end{thebibliography}

\end{document}